\documentclass[sigconf]{acmart}

\usepackage{booktabs} 
\usepackage{todonotes}
\usepackage{multicol}
\usepackage{amsmath}
\usepackage{xcolor}

\newcommand{\eat}[1]{}
\newcommand{\todoKathy}[1]{\todo[inline,color=green!30!white]{Kathy: #1}}

\renewcommand{\AA}{\mathcal{A}}

\newcommand{\OO}{\mathcal{O}}

\renewcommand{\SS}{\mathcal{S}}

\setcopyright{acmcopyright}

\begin{document}

\copyrightyear{2019} 
\acmYear{2019} 
\setcopyright{acmcopyright}
\acmConference[ICCPS '19]{10th ACM/IEEE International Conference on Cyber-Physical Systems (with CPS-IoT Week 2019)}{April 16--18, 2019}{Montreal, QC, Canada}
\acmBooktitle{10th ACM/IEEE International Conference on Cyber-Physical Systems (with CPS-IoT Week 2019) (ICCPS '19), April 16--18, 2019, Montreal, QC, Canada}
\acmPrice{15.00}
\acmDOI{10.1145/3302509.3313784}
\acmISBN{978-1-4503-6285-6/19/04}

\title[Simulation to Scaled City: Zero-Shot Policy Transfer for Traffic Control]{Simulation to Scaled City: Zero-Shot Policy Transfer for Traffic Control via Autonomous Vehicles}

\author{Kathy Jang}
\affiliation{%
  \institution{University of California, Berkeley}
}
\email{kathyjang@berkeley.edu}

\author{Eugene Vinitsky}
\affiliation{%
\institution{University of California, Berkeley}
}
\email{evinitsky@berkeley.edu}

\author{Behdad Chalaki}
\affiliation{%
  \institution{University of Delaware}
}
\email{bchalaki@udel.edu}

 \author{Ben Remer}
\affiliation{%
  \institution{University of Delaware}
  }
\email{bremer@udel.edu}

\author{Logan Beaver}
\affiliation{%
  \institution{University of Delaware}
}
\email{lebeaver@udel.edu}

\author{Andreas A. Malikopoulos}
\affiliation{%
  \institution{University of Delaware}
}
\email{andreas@udel.edu}

\author{Alexandre Bayen}
\affiliation{%
  \institution{Univesity of California, Berkeley}
}
\email{bayen@berkeley.edu}

\renewcommand{\shortauthors}{K. Jang et al.}

\begin{abstract}
Using deep reinforcement learning, we successfully train a set of two autonomous vehicles to lead a fleet of vehicles onto a roundabout and then transfer this policy from simulation to a scaled city without fine-tuning. We use \emph{Flow}, a library for deep reinforcement learning in microsimulators, to train two policies, (1) a policy with noise injected into the state and action space and (2) a policy without any injected noise. In simulation, the autonomous vehicles learn an emergent metering behavior for both policies which allows smooth merging. We then directly transfer this policy without any tuning to the \emph{University of Delaware's Scaled Smart City (UDSSC)}, a 1:25 scale testbed for connected and automated vehicles.  We characterize the performance of the transferred policy based on how thoroughly the ramp metering behavior is captured in UDSSC. We show that the noise-free policy results in severe slowdowns and  only, occasionally, it exhibits acceptable metering behavior. On the other hand, the noise-injected policy consistently performs an acceptable metering behavior, implying that the noise eventually aids with the zero-shot policy transfer. Finally, the transferred, noise-injected policy leads to a $5\%$ reduction of average travel time and a reduction of $22\%$ in maximum travel time in the UDSSC.       Videos of the proposed self-learning controllers can be found at \textcolor{blue}{\url{https://sites.google.com/view/iccps-policy-transfer}}.
\end{abstract}



\maketitle

\section{Introduction}
\textbf{Control of mixed-autonomy traffic:}
Transportation is a major source of US energy consumption and greenhouse gas emissions, accounting for 28\% and 26\% respectively. According to the bureau of transportation statistics, total road miles traveled is continuously increasing, growing at 2 to 3\% per year between 2010 and 2014 while over the same period the total road length of the US transportation network remained unchanged. The increased road usage is coupled with an increase in congestion. Overall congestion delay in 2014 was 6.9 billion hours, an increase of 33\% since 2000; the problem is even worse in metropolitan areas where travelers needed to allocate an additional 150\% more travel time during peak periods to arrive on time. The congestion also has significant economic cost, totaling 160 billion dollars in 2014 \cite{US2016}. 
Depending on their usage, automated vehicles have the potential to alleviate system level metrics such as \emph{congestion}, \emph{accident rates}, and \emph{greenhouse gas emissions} through a combination of intelligent routing, smoother driving behavior, and faster reaction time \cite{Wadud2016}.

Partially automated systems are predicted to increasingly populate roadways between 2020 and 2025 but will primarily be usable in high driving or high speed operations in light traffic. Hazard detection technology is not expected to be mature enough for full automation in the presence of general vehicles and pedestrians (i.e. heterogeneous fleets, manned/unmanned, bicycles, pedestrians, mixed use road-space etc.) until at least 2030. It takes 20 years for a vehicle fleet to turn over sufficiently which makes it likely that vehicles will be partially manned at least until 2050~\cite{shladover2017connected}.

Recently, the steady increase in usage of cruise control systems on the roadway offers an opportunity to study the optimization of traffic in the framework of \emph{mixed-autonomy traffic}: traffic that is partially automated but mostly still consists of human driven vehicles. However, the control problems posed in this framework are notoriously difficult to solve. Traffic problems, which often exhibit features such as time-delay, non-linear dynamics, and hybrid behavior, are challenging for classical control approaches, as microscopic traffic models are high complexity: discrete events (lane changes, traffic light switches), continuous states (position, speed, acceleration), and non-linear driving models. These complexities make analytical solutions often intractable. The variety and non-linearity of traffic often leads to difficult trade-offs between the fidelity of the dynamics model and tractability of the approach. 

\textbf{Classical control approaches:}
Classical control approaches have successfully solved situations in which the complexity of the problem can be reduced without throwing away key aspects of the dynamics. For example, there is a variety of analytical work on control of autonomous intersections with simple geometries. For mixed-autonomy problems, there have been significant classical controls based results for simple scenarios like vehicles on a ring~\cite{cui2017stabilizing} or a single lane of traffic whose stability can be characterized~\cite{orosz2016connected,swaroop1996string}. A thorough literature review on coordinating autonomous vehicles in intersections, merging roadways, and roundabouts can be found in \cite{Rios-Torres2017}. The classical control approaches described in this review can be broken down into reservation methods, scheduling, optimization with safety constraints, and safety maximization. Other approaches discussed in the review involve applications of queuing theory, game theory, and mechanism design.

However, as the complexity of the problem statement increases, classical techniques become increasingly difficult to apply. Shifting focus from simple scenarios to, for example, hybrid systems with coexisting continuous and discrete controllers, explicit guarantees for hand-designed controllers can become harder to find. Ultimately, when the complexity of the problem becomes too high, optimization-based approaches have been shown to be a successful approach in a wide variety of domains from robotics~\cite{kuindersma2016optimization} to control of transportation infrastructure~\cite{li2016traffic}.

\textbf{Deep reinforcement learning:} \ \textit{Deep reinforcement learning} (deep RL) has recently emerged as an effective technique for control in high dimensional, complex CPS systems. Deep RL has shown promise for the control of complex, unstructured problems as varied as robotic skills learning~\cite{gu2017deep}, playing games such as Go~\cite{silver2017mastering}, and traffic light ramp metering~\cite{belletti2017expert}.
Of particular relevance to this work, deep RL has been successful in training a single autonomous vehicle to optimize traffic flow in the presence of human drivers~\cite{wu2017flow}.

One key distinction in RL is whether the algorithm is model-free or model-based, referring to whether the algorithm is able to query a dynamics model in the computation of the control or the policy update. Model-free RL tends to outperform model-based RL if given sufficient optimization time, but requires longer training times. Thus, model-free techniques are most effective when samples can be cheaply and rapidly generated. This often means that model-free RL works best in simulated settings where a simulation step can be made faster than real-time and simulation can be distributed across multiple CPUs or GPUs. A long-standing goal is to be able to train a controller in simulation, where model-free techniques can be used, and then use the trained controller to control the actual system. 

\textbf{Policy Transfer:}
Transfer of a controller from a training domain to a new domain is referred to as \emph{policy transfer}. 
The case where the policy is directly transferred without any fine-tuning is referred to as \emph{zero-shot policy transfer.} Zero-shot policy transfer is a difficult problem in RL, as the true dynamics of the system may be quite different from the simulated dynamics, an issue referred to as \emph{model mismatch}. Techniques used to overcome this include adversarial training, in which the policy is trained in the presence of an adversary that can modify the dynamics and controller outputs and the policy must subsequently become robust to perturbations~\cite{pinto2017robust}. Other techniques to overcome \emph{model mismatch} include re-learning a portion of the controller\cite{rusu2016sim}, adding noise to the dynamics model~\cite{peng2017sim}, and learning a model of the true dynamics that can be used to correctly execute the desired trajectory of the simulation-trained controller~\cite{christiano2016transfer}. Efforts to overcome the reality gap have been explored in vision-based reinforcement learning~\cite{muller2018transfer} and in single AV systems~\cite{xu2018transfer}.

Other challenges with policy transfer include \emph{domain mismatch}, where the true environment contains states that are unobserved or different from simulation. For example, an autonomous vehicle might see a car color that is unobserved in its simulations and subsequently react incorrectly. Essentially, the controller overfits to its observed states and does not generalize. Domain mismatch can also occur as a result of imperfect sensing or discrepancies between the simulation and deployment environment. While in simulation it is possible to obtain perfect observations, this is not always the case in the real world. Small differences between domains can lead to drastic differences in output. For example, a slight geometric difference between simulation and real world could result in a vehicle being registered as being on one road segment, when it is on another. This could affect the control scheme in a number of ways, such as a premature traffic light phase change. Techniques used to tackle this problem include domain randomization~\cite{tobin2017domain}, in which noise is injected into the state space to enforce robustness with respect to unobserved states. 

\textbf{Contributions and organization of the article:} 
In this work we use deep RL to train two autonomous vehicles to learn a classic form of control: ramp metering, in which traffic flow is regulated such that one flow of vehicles is slowed such that another flow can travel faster. While in the real world, ramp metering is controlled via metering lights, we demonstrate the same behavior using AVs instead of lights. Each RL vehicle interacts with sensors at each of the entrance ramps and is additionally able to acquire state information about vehicles on the roundabout, as well as state information about the other RL vehicle. By incorporating this additional sensor information, we attempt to learn a policy that can time the merges of the RL vehicles and their platoons to learn ramp metering behavior, which prevents energy-inefficient decelerations and accelerations. Being positioned at the front of a platoon of vehicles, each RL vehicle has the ability to control the behavior of the platoon of human-driven vehicles following it. The RL vehicle, also referred to in this paper as an autonomous vehicle (AV), is trained with the goal of minimizing the average delay of all the vehicles in simulation. 

Next, we show how we overcome the RL to real world reality gap and demonstrate RL's real world relevance by transferring the controllers to the University of Delaware's Scaled Smart City (UDSSC), a reduced-scale city whose dynamics, which include sensor delays, friction, and actuation, are likely closer to true vehicle dynamics. RL trained policies, which are learned in a simulation environment, can overfit to the dynamics and observed states of the simulator and can then fare poorly when transferred to the real world. The combination of model and domain mismatch contributes to this problem. We combine the ideas of domain randomization with adversarial perturbations to the dynamics and train a controller in the presence of noise in both its observations and actions. For reasons discussed in Sec. ~\ref{sec:discussion}, we expect the addition of noise in both state and action to help account for both model and domain mismatch.

In this work we present the following results:
\begin{itemize}
    \item The use of deep RL in simulation to learn an emergent metering policy.
    \item A demonstration that direct policy transfer to UDSSC leads to poor performance.
    \item A successful zero-shot policy transfer of the simulated policy to the UDSSC vehicles via injection of noise into both the state and action space. 
    \item An analysis of the improvements that the autonomous vehicles bring to the congested roundabout.
\end{itemize}

The remainder of the article is organized as follows:
\begin{enumerate}
    \item Section \ref{sec:background} provides an introduction to deep RL and the algorithms used in this work.
    \item Section \ref{sec:experiments} describes the setup we use to learn control policies via RL, followed by the policy transfer process from simulation to the physical world.
    \item Section \ref{sec:discussion} discusses the results of our experiments and provides intuition for the effectiveness of the state and action noise.
    \item Section \ref{sec:conclusions} summarizes our work and future directions.
\end{enumerate}

\section{BACKGROUND}
\label{sec:background}
\subsection{Reinforcement Learning}
In this section, we discuss the notation and briefly describe the key concepts used in RL. RL focuses on deriving optimal controllers for  \textit{Markov decision processes}  (MDP)~\cite{bellman1957markovian}.
The system described in this article solves tasks which conform to the standard structure of a finite-horizon discounted MDP, defined by the tuple $(\mathcal{S}, \mathcal{A}, P, r, \rho_0, \gamma, T)$. Here $\mathcal{S}$ is a  set of states and $\mathcal{A}$ is a set of actions where both sets can be finite or infinite. $P: \mathcal{S} \times \mathcal{A} \times \mathcal{S} \to \mathbb{R}_{\geq 0}$ is the transition probability distribution describing the probability of moving from one state $s$ to another state $s'$ given action $a$, $r : \SS \times \AA \to \mathbb{R}$ is the reward function, $\rho_0: \SS \to \mathbb{R}_{\geq 0}$ is the probability distribution over start states, $\gamma \in (0, 1]$ is the discount factor, and $T$ is the horizon. For partially observable tasks, which conform to the structure of a \textit{partially observable Markov decision process} (POMDP), two more components are required, namely $\Omega$, a set of observations of the hidden states, and $\OO: \SS \times \Omega \to \mathbb{R}_{\geq 0}$, the observation probability distribution.


RL studies the problem of how an agent can learn to take actions in its environment to maximize its expected cumulative discounted reward: specifically it tries to optimize $R = \mathbb{E} \left[\sum_{t=0}^T \gamma^t r_t \right]$ where $r_t$ is the reward at time $t$. The goal is to use the observed data from the MDP to optimize a \emph{policy} $\Pi: \SS \to \AA$, mapping states to actions, that maximizes $R$. This policy can be viewed as the controller for the system, however, we stick to the convention of RL literature and refer to it as a policy. It is increasingly common to parameterize the policy via a neural net. We will denote the parameters of this policy, which are the weights of the neural network, by $\theta$ and the policy by $\pi_\theta$. A neural net consists of a stacked set of affine linear transforms and non-linearities that the input is alternately passed through. The presence of multiple stacked layers is the origin of the term "deep"-RL.

\subsection{Policy Gradient Methods}
Policy gradient methods use Monte Carlo estimation to compute an estimate of the gradient of the expected discounted reward $\nabla_\theta R = \nabla_\theta \mathbb{E}\left[\sum_{t=0}^T \gamma^t r_t \right]$ where $\theta$ are the parameters of the policy $\pi_\theta$. We perform repeated \emph{rollouts}, in which the policy is used to generate the actions at each time step. At the end of the rollout, we have accumulated a state, action, reward trajectory $\tau = (s_0, a_0, r_0, \dots, s_T)$. Policy gradient methods take in a set of these trajectories and use them to compute an estimate of the gradient $\nabla_\theta R$ which can be used in any gradient ascent-type method.

The particular policy gradient method used in this paper is \emph{Trust Region Policy Optimization} (TRPO) ~\cite{schulman2015trust}. TRPO is a monotonic policy improvement algorithm, whose update step provides guarantees of an increase in the expected total reward. However, the exact expression for the policy update leads to excessively small steps so implementations of TRPO take larger steps by using a trust region. In this case, the trust region is a bound on the KL divergence between the old policy and the policy update. While not a true distance measure, a small KL divergence between the two policies suggests that the policies do not act too differently over the observed set of states, preventing the policy update step from sharply shifting the policy behavior. 

\subsection{Car Following Models}
\label{sec:car-following}
For our model of the driving dynamics, we used the \textit{Intelligent Driver Model}~\cite{Treiber2000} (IDM) that is built into the traffic microsimulator SUMO~\cite{SUMO2012}.
IDM is a microscopic car-following model commonly used to model realistic driver behavior. Using this model, the acceleration for vehicle $\alpha$ is determined by its bumper-to-bumper \emph{headway} $s_\alpha$ (distance to preceding vehicle), the vehicle's own velocity $v_\alpha$, and relative velocity $\Delta v_\alpha$, via the following equation:
\begin{equation} \label{eq:idm}
a_{\text{IDM}} = \frac{dv_\alpha}{dt} = a \bigg[ 1 - \bigg( \frac{v_\alpha}{v_0} \bigg)^\delta - \bigg( \frac{s^*(v_\alpha,\Delta v_\alpha)}{s_\alpha} \bigg)^2 \bigg]
\end{equation}
where $s^*$ is the desired headway of the vehicle, denoted by:
\begin{equation} \label{eq:s_star}
s^*(v_\alpha,\Delta v_\alpha) = s_0 + \max \bigg( 0, v_\alpha T + \frac{v_\alpha \Delta v_\alpha}{2 \sqrt{ab}} \bigg)
\end{equation}
where $s_0, v_0, T, \delta, a, b$ are given parameters. Typical values for these parameters can be found in ~\cite{Treiber2000}; the values used in our simulations are given in Sec.~\ref{sec:FLOW-setup}.  To better model the natural variability in driving behavior, we induce stochasticity in the desired driving speed $v_0$. For a given vehicle, the value of $v_0$ is sampled from a Gaussian whose mean is the speed limit of the lane and whose standard deviation is 20\% of the speed limit. 

Car following models are not inherently collision-free, we supplement them with a safe following rule: a vehicle is not allowed to take on velocity values that might lead to a crash if its lead vehicle starts braking at maximum deceleration. However, due to some uncertainty in merging behavior, there are still rare crashes that can occur in the system. 

\subsection{Flow}
\label{sec:Flow}
We run our experiments in \emph{Flow}~\cite{wu2017flow}, a library that provides an interface between a traffic microsimulator, SUMO~\cite{SUMO2012}, and two RL libraries, rllab~\cite{duan2016benchmarking} and RLlib~\cite{liang2017ray}, which are centralized and distributed RL libraries respectively. \emph{Flow} enables users to create new traffic networks via a Python interface, introduce autonomous controllers into the networks, and then train the controllers in a distributed system on the cloud via AWS EC2. To make it easier to reproduce our experiments or try to improve on our benchmarks, the code for \emph{Flow}, scripts for running our experiments, and tutorials can be found at \textcolor{blue}{\url{https://github.com/flow-project/flow}}.

\begin{figure}
\centering
\includegraphics[width=0.4\textwidth]{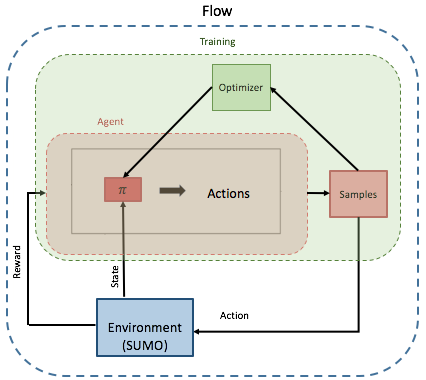}
\caption{Diagram of the iterative process in \emph{Flow}. Portions in red correspond to the controller and rollout process, green to the training process, and blue to traffic simulation.}
\label{fig:flow}
\end{figure}
Fig.~\ref{fig:flow} describes the process of training the policy in \emph{Flow}. The controller, here represented by policy $\pi_\theta$, receives a state and reward from the environment and uses the state to compute an action. The action is taken in by the traffic microsimulator, which outputs the next state and a reward. The (state, next state, action, reward) tuple are stored as a sample to be used in the optimization step. After accumulating enough samples, the states, actions, and rewards are passed to the optimizer to compute a new policy. 

\subsection{University of Delaware's Scaled Smart City (UDSSC)}
\begin{figure}
\centering
\includegraphics[width=0.45\textwidth]{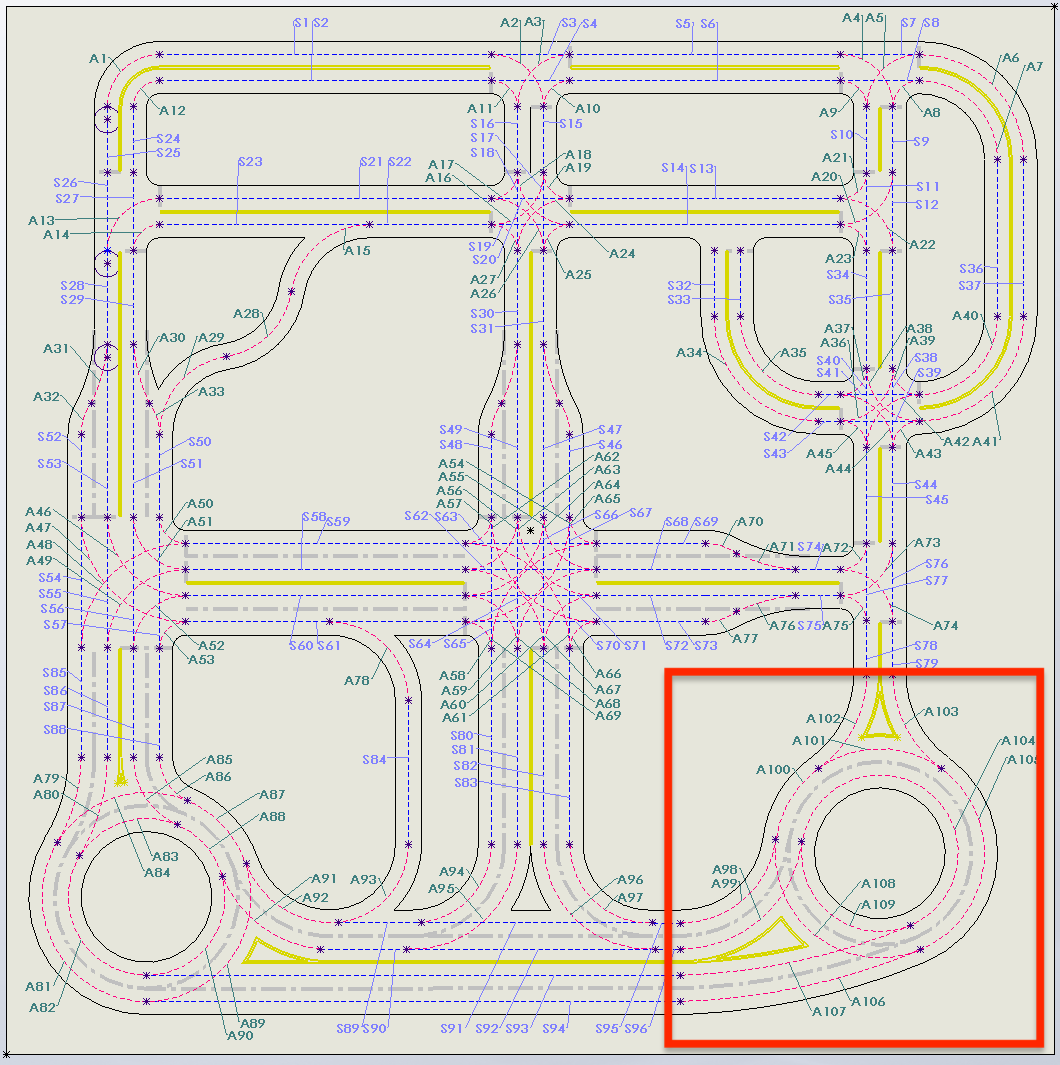}
\caption{Diagram of the UDSSC road map, with the experimental zone highlighted in red.}
\label{fig:UDSSC_map}
\end{figure}

The \textit{University of Delaware's Scaled Smart City (UDSSC)} was used to validate the performance of the RL control system. 
UDSSC is a testbed (1:25 scale) that can help prove concepts beyond the simulation level and can replicate real-world traffic scenarios in a small and controlled environment. UDSSC uses a VICON camera system to track the position of each vehicle with sub-millimeter accuracy, which is used both for control and data collection. The controller for each vehicle is offloaded to a mainframe computer and runs on an independent thread which is continuously fed data from the VICON system. Each controller uses the global VICON data to generate a speed reference for the vehicles allowing for precise independent closed-loop feedback control. A detailed description of UDSSC can be found in \cite{Stager2018}. To validate the effectiveness of the proposed RL approach in a physical environment, the southeast roundabout of the UDSSC was used (Fig. \ref{fig:UDSSC_map}).

\section{EXPERIMENTAL DEPLOYMENT}
\label{sec:experiments}
\subsection{Experimental setup}
\subsubsection{Simulation Details}
\label{sec:simulation-details}
\begin{figure}
\centering
\includegraphics[width=0.45\textwidth]{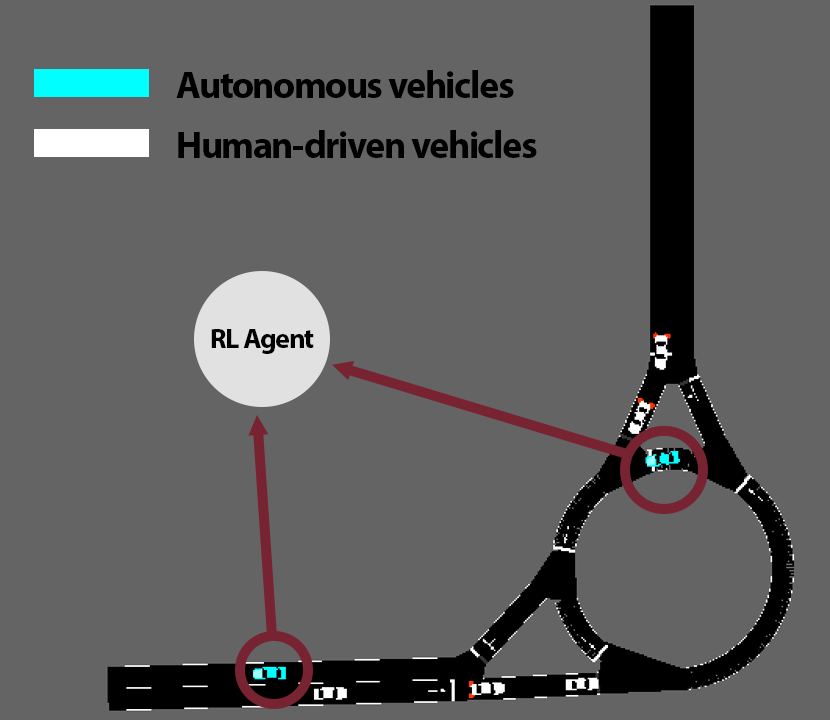} 
\caption{SUMO-generated network in UDSSC's roundabout. The blue vehicles are the AVs; they are both controlled by the RL policy. Videos of this policy in simulation are available at \textcolor{blue}{\url{https://sites.google.com/view/iccps-policy-transfer}}.}
\label{fig:roundabout}
\end{figure}

To derive the RL policy, we developed a model of the roundabout highlighted in red in Fig.~\ref{fig:UDSSC_map} in SUMO. The training of the model, shown in Fig.~\ref{fig:roundabout}, included a single-lane roundabout with entry points at the northern and western ends. Throughout this paper we will refer to vehicles entering from the western end as the \emph{western platoon} and the north entrance as the \emph{northern platoon}. The entry points of the model are angled slightly different as can be seen in Figs.~\ref{fig:UDSSC_map} and \ref{fig:roundabout}.

The human-controlled vehicles operate using SUMO's built-in IDM controller, with several modified parameters. In these experiments, the vehicles operating with the IDM controller are run with $T=1$, $a=1$, $b=1.5$, $\delta=4$, $s_0=2$, $v_0=30$, and noise$\:=0.1$, where $T$ is a safe time headway, $a$ is a comfortable acceleration in $m/s^2$, $b$ is a comfortable deceleration, $\delta$ is an acceleration exponent, $s0$ is the linear jam distance, $v_0$ is a desired driving velocity, and noise is the standard deviation of a zero-mean normal perturbation to the acceleration or deceleration. Details of the physical interpretation of these parameters can be found in \cite{Treiber2000}. 
Environment parameters in simulation were set to match the physical constraints of UDSSC. These include: a maximum acceleration of $1 \frac{\text{m}}{\text{s}^2}$, a maximum deceleration of $-1 \frac{\text{m}}{\text{s}^2}$, and a maximum velocity of $15  \frac{\text{m}}{\text{s}}$. The timestep of the system is set to $1.0$ seconds. 

Simulations on this scenario in the roundabout were executed across a range of different settings in terms of volume and stochasticity of inflows. In the RL policy implemented in UDSSC and discussed in \ref{sec:UDSCC-setup}, vehicles are introduced to the system via deterministic inflows from the northern and western ends of the roundabout using two routes: (1) the northern platoon enters the system from the northern inflow, merges into the roundabout, and exits through the western outflow and (2) the western platoon enters the system from the western inflow, U-turns through the roundabout, and exits through the western outflow. The western platoon consists of four vehicles total: three vehicles controlled with the IDM controller led by a vehicle running with the RL policy. The northern platoon consists of of three vehicles total: two vehicles controlled with the IDM controller led by a vehicle running with the RL policy. New platoons enter the system every 1.2 minutes, the rate of which is significantly sped up in simulation. These inflow settings are designed to showcase the scenario where routes clash (Fig. \ref{fig:roundabout}).


\label{sec:FLOW-setup}
\subsubsection{UDSSC}
\label{sec:UDSCC-setup}

\begin{figure}
\centering
\includegraphics[width=0.45\textwidth]{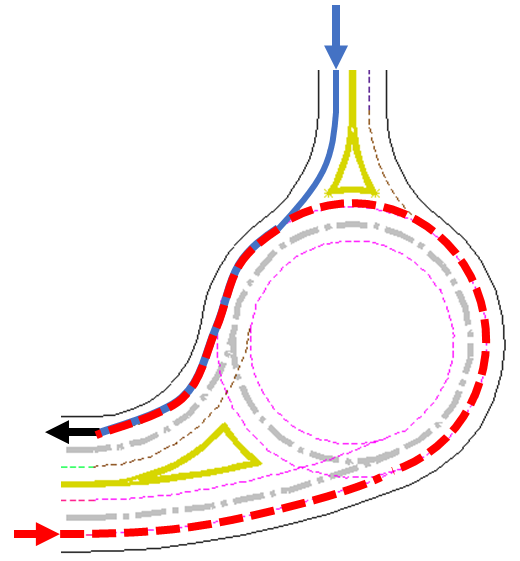}
\caption{Visualization of the path taken on the UDSSC roundabout. The red route enters going east and exits going west; the blue route enters north and exits via the same west entrance as the red route. }
\label{fig:UDSSC-loop}
\end{figure}




Each vehicle in UDSSC uses a saturated IDM controller to (1) avoid negative speeds, (2) ensure that the rear-end collision constraints do not become active, and (3) maintain the behavior of Eq.~(\ref{eq:idm}). Both the IDM and RL controllers provide a desired acceleration for the vehicles, which is numerically integrated to calculate each vehicle's reference speed.

Merging at the northern entrance of the roundabout is achieved by an appropriate yielding function. Using this function, the car entering the roundabout proceeds only if no other vehicle is on the roundabout at a distance from which a potential lateral collision may occur. Otherwise, the vehicle stops at the entry of the roundabout waiting to find a safe space to proceed.


To match the SUMO training environment, only one vehicle per path was allowed to use the RL policy. The paths taken by each vehicle are shown in Fig. \ref{fig:UDSSC-loop}. For each path, the first vehicle to enter the experimental zone was controlled by the RL policy; every subsequent vehicle runs with the saturated IDM controller. Once an active vehicle running with the RL policy exits the experimental zone, it reverts back to the IDM controller.

In the experiments, the vehicles operated in a predefined deterministic order, as described in \ref{sec:simulation-details}. Four vehicles were placed just outside the experimental zone on the western loop, and three vehicles were placed in the same fashion near the northern entrance. Each vehicle platoon was led by a vehicle running with the RL policy, except in the baseline case where all vehicles used the saturated IDM controller. The experiment was executed with three variations: (1) the baseline case with all vehicles running with the IDM controller, (2) the case with a leader vehicle running with an RL policy trained in SUMO, and (3) the case where the leader vehicles running with an RL policy were trained in simulation with noise injected into their observations and accelerations.



\subsection{Reinforcement Learning Structure}
\label{sec:VSLParam}
\subsubsection{Action space}
We parametrize the controller as a neural net mapping the observations to a mean and diagonal covariance matrix of a Gaussian. The actions are sampled from the Gaussian; this is a standard controller parametrization~\cite{levine2014learning}. 
The actions are a two-dimensional vector of accelerations in which the first element corresponds to vehicles on the north route and the second element to the west route. Because the dimension of the action vector is fixed, there can only ever be 1 AV from the northern entry and 1 AV from the western entry. Two queues, one for either entryway, maintain a list of the RL-capable vehicles that are currently in the system. It should be noted that the inflow rates are chosen such that the trained policy never contains more than a queue of length 2. Platoons are given ample time to enter and exit the system before the next platoon arrives. This queue mechanism is designed to support the earlier stages of training, when RL vehicles are learning how to drive, which can result in multiple sets of platoons and thus more than 2 RL vehicles being in the system at the same time. Control is given to vehicles at the front of both queues. When a vehicle completes its route and exits the experimental zone, its ID is popped from the queue. All other RL-capable vehicles are passed IDM actions until they reach the front of the queue. If there are fewer than two AVs in the system, the extra actions are simply unused.

The dynamics model of the autonomous vehicles are  given by the IDM described in sec. \ref{sec:car-following} subject to a minimum and maximum speed i.e.
\begin{equation}
v^{\text{IDM}}_j(t+\Delta t) = \text{max}\left(\text{min}\left(v_{AV}(t) + a_{IDM}\Delta t, v^{\text{max}}_j(t)\right), 0\right)
\end{equation}
where $v^{AV}_j(t)$ is the velocity of autonomous vehicle $j$ at time $t$, $a_{IDM}$ is the acceleration given by an IDM controller, $\Delta t$ is the time-step, and $v^{\text{max}}_j(t)$ is the maximum speed set by the city $j$. For the AVs, the acceleration $a_t$ is straightforwardly added to the velocity via a first-order Euler integration step 
\begin{equation}
v^{\text{AV}}_j(t+\Delta t) = \text{max}\left( \text{min}\left(v^{\text{AV}}_j(t) + a_{t}\Delta t, v^{\text{max}}_j(t)\right), 0\right)
\end{equation}

\subsubsection{Observation space}
For the purposes of keeping in mind physical sensing constraints, the state space of the MDP is partially observable. It is normalized to $\pm 1$ and includes the following:
\begin{itemize}
\item The positions of the AVs.
\item The velocities of the AVs.
\item The distances from the roundabout of the 6 closest vehicles to the roundabout for both roundabout entryways.
\item The velocities of the 6 closest vehicles to the roundabout for both roundabout entryways.
\item Tailway and headway (i.e. distances to the leading and following vehicles) of vehicles from both AVs.
\item Length of the number of vehicles waiting to enter the roundabout for both roundabout entryways.
\item The distances and velocities of all vehicles in the roundabout.
\end{itemize}
This state space was designed with real-world implementation in mind, and could conceivably be implemented on existing roadways equipped with loop detectors, sensing tubes, and vehicle-to-vehicle communication between the AVs. For a sufficiently small roundabout, it is possible that an AV equipped with enough cameras could identify the relevant positions and velocities of roundabout vehicles. Similarly, the queue lengths can be accomplished with loop detectors, and the local information of the AVs (its own position and velocity, as well as the position and velocity of its leader and follower) are already necessarily implemented in distance-keeping cruise control systems. 

\subsubsection{Action and State Noise}
The action and state spaces are where we introduce noise with the purpose of training a more generalizable policy that is more resistant to the difficulties of cross-domain transfer. We train the policies in two scenarios, a scenario where both the action and state space are perturbed with noise and a scenario with no noise. This former setting corresponds to a type of \emph{domain randomization}. In the noisy case, we draw unique perturbations for each element of the action and state space from a Gaussian distribution with zero mean and a standard deviation of $0.1$. In the action space, which is composed of just accelerations, this corresponds to a standard deviation of $0.1 \frac{\text{m}}{\text{s}^2}$. In the state space, which is normalized to 1, this corresponds to a standard deviation of 1.5 m/s for velocity-based measures. The real-life deviations of each distance-based state space element are described here: AV positions, and the tailways and headways of the AVs, deviate the most at 44.3 m, the large uncertainty of which results in a policy that plays it safe. The distance from the northern and western entryways respectively deviate by 7.43m and 8.66 m. The length of the number of vehicles waiting to enter the roundabout from the northern and western entryway respectively deviate by 1.6 and 1.9 vehicles.

These perturbations are added to each element of the action and state space.
The elements of the action space are clipped to the  maximum acceleration and deceleration of $\pm 1$, while the elements of the state space are clipped to $\pm 1$ to maintain normalized boundaries. 
Noisy action and state spaces introduce uncertainty to the training process. The trained policy must still be effective even in the presence of uncertainty in its state as well as uncertainty that its requested actions will be faithfully implemented.

\subsubsection{Reward function}
For our reward function we use a combination of the L2-norm of the velocity of all vehicles in the system and penalties discouraging standstills or low velocity travel. 
\begin{equation}
r_t = \frac{\max\left({v_\text{max}\sqrt{n} - \sqrt{\sum_{i=1}^{n}(v_{i, t}-v_\text{max})^2}}, 0\right)}{v_\text{max}\sqrt{n}} -
1.5\cdot \text{pen}_\text{s} - \text{pen}_\text{p}
\end{equation}
where $n$ is the number of all vehicles in the system, $v_\text{max}$ is the maximum velocity of $15 \frac{\text{m}}{\text{s}}$, $v_{i, t}$ is the velocity that vehicle $v_i$ is travelling at at time $t$. The first term incentives vehicles to travel near speed $v_\text{max}$ but also encourages the system to prefer a mixture of low and high velocities versus a mixture of mostly equal velocities. The preference for low and high velocities is intended to induce a platooning behavior. RL algorithms are sensitive to the scale of the reward functions; to remove this effect the reward is
normalized by $v_\text{max} \sqrt{n}$ so that the maximum reward of a time-step is 1. 

This reward function also introduces 2 penalty functions, $\text{pen}_s$ and $\text{pen}_p$. $\text{pen}_s$ returns the number of vehicles that are traveling at a velocity of 0, and $\text{pen}_p$ is the number of vehicles that are traveling below a velocity of 0.3 m/s. They are defined as:
\begin{equation}
    \text{pen}_\text{s} = \sum_{i=1}^n g(i) \ \
    \text{where} \ \ \
    g(x) = \begin{cases}
             0, & v_x \neq 0,\\
             1, & v_x = 0.
            \end{cases}
\end{equation}
\begin{equation}
    \text{pen}_\text{p} = \sum_{i=1}^n{h(i)} \ \ 
    \text{where} \ \ \ 
    h(x) = \begin{cases}
             0, & v_x \geq 0.3,\\
             1, & v_x \leq 0.3
            \end{cases}
\end{equation}
These penalty functions are added to discourage the autonomous vehicle from fully stopping or adopting near-zero speeds. In the absence of these rewards, the RL policy learns to game the simulator by blocking vehicles from entering the simulator on one of the routes, which allows for extremely high velocities on the other route. This occurs because velocities of vehicles that have not yet emerged from an inflow do not register, so no penalties are incurred when the AV blocks further vehicles from entering the inflow.

\subsection{Algorithm/simulation details}
We ran the RL experiments with a discount factor of $.999$, a trust-region size of $.01$, a batch size of $20000$, a horizon of $500$ seconds, and trained over $100$ iterations. The controller is a neural network, a \emph{Gaussian multi-layer perceptron} (MLP), with hidden sizes of $(100, 50, 25)$ and a tanh non-linearity. The choice of neural network non-linearities, size, and type were picked based on traffic controllers developed in \cite{vinitsky2018benchmarks}. The states are normalized so that they are between $0$ and $1$ by dividing each states by its maximum possible value. The actions are clipped to be between $-1$ and $1$. Both normalization and clipping occur after the noise is added to the system so that the bounds are properly respected. 

\subsection{Code reproducibility}
In line with open, reproducible science, the following codebases are needed to reproduce the results of our work. \emph{Flow} can be found at \textcolor{blue}{\url{https://github.com/flow-project/flow}}. The version of \emph{rllab} used for the RL algorithms is available at \textcolor{blue}{\url{https://github.com/cathywu/rllab-multiagent}} at commit number \textbf{4b5758f}. \emph{SUMO} can be found at \textcolor{blue}{\url{https://github.com/eclipse/sumo}} at commit number \textbf{1d4338ab80}.

\subsection{Policy Transfer}

The RL policy learned through Flow was encoded as the weights of a neural network. These weights were extracted from a serialized file and accessed via a Python function which maps inputs and outputs identical to those used in training. Separating these weights from rllab enables an interface for state space information from the UDSSC to be piped straight into the Python function, returning the accelerations to be used on the UDSSC vehicles. The Python function behaves as a control module within the UDSSC, replacing the IDM control module in vehicles operating under the RL policy.

The inputs to the RL neural network were captured by the VICON system and mainframe. The global 2D positions of each vehicle were captured at each time step. These positions were numerically derived to get each vehicle's speed and were compared to the physical bounds on the roadways to get the number of vehicles in each queue at the entry points. Finally, the 2D positions were mapped into the 1D absolute coordinate frame used during training. This array was passed into the RL control module as the inputs of the neural network.

\subsection{Results}
\label{sec:results}
\begin{figure}
\centering
\includegraphics[width=0.45\textwidth]{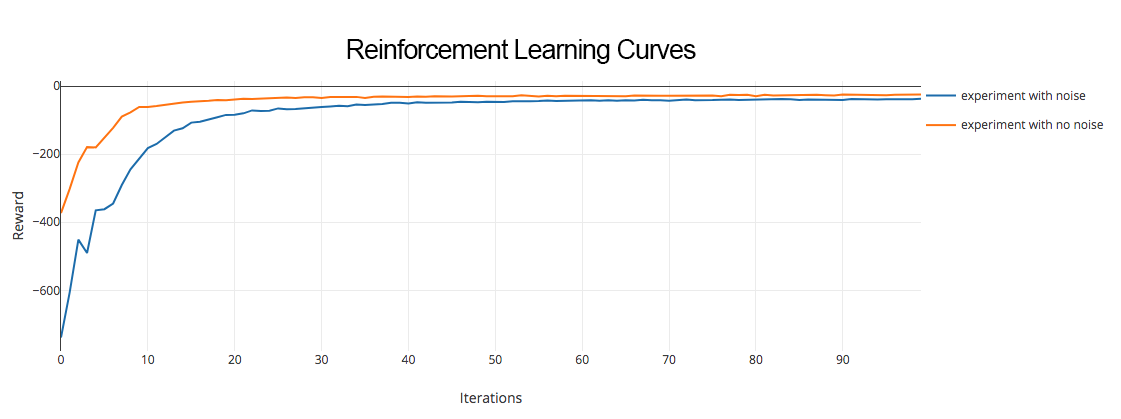}
\caption{Convergence of the RL reward curve of an experiment with noised IDM, RL accelerations, and noisy state space}
\label{fig:reward}
\end{figure}

\begin{figure}
    \centering
    \includegraphics[width=0.45\textwidth]{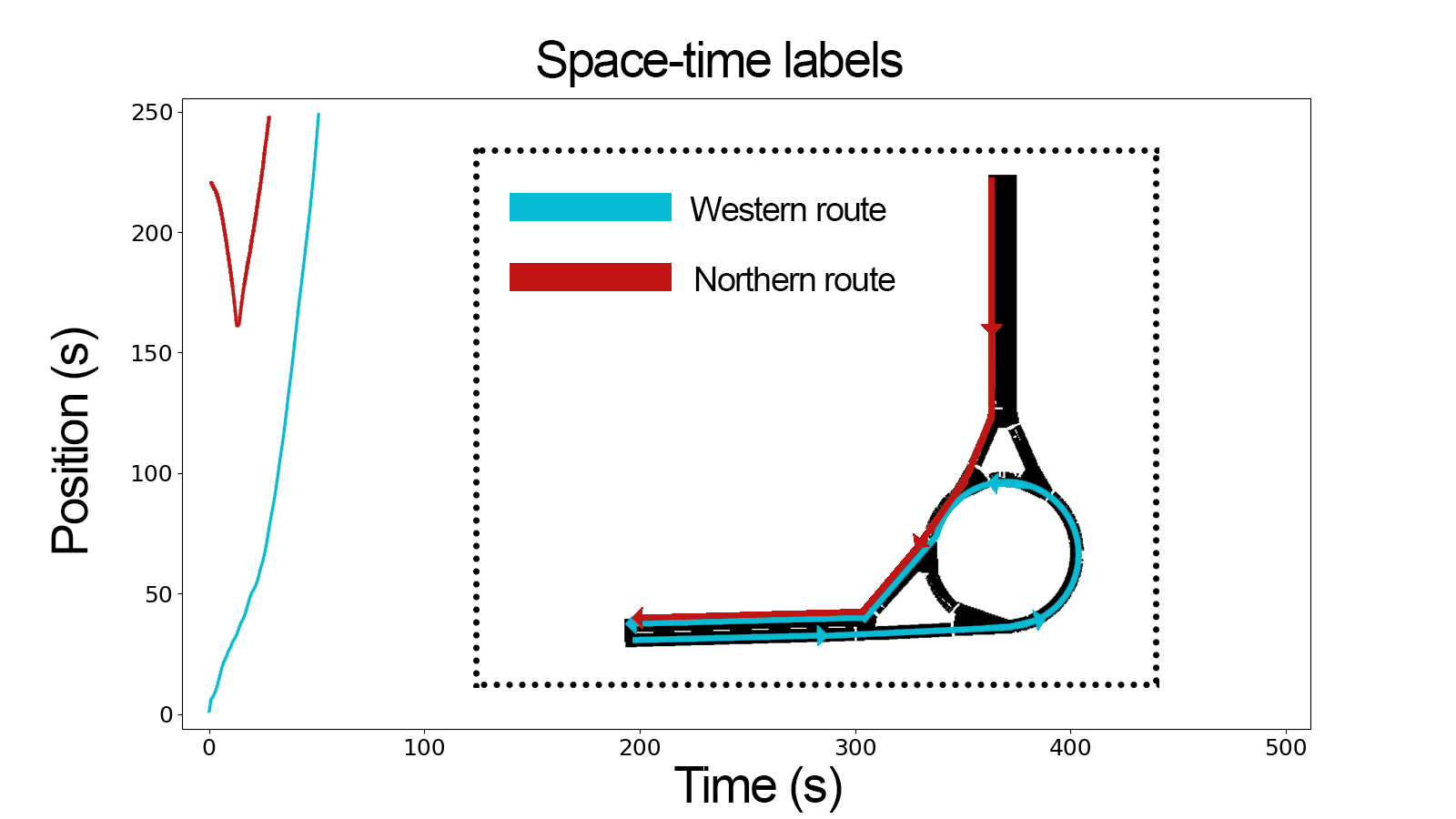}
    \includegraphics[width=0.45\textwidth]{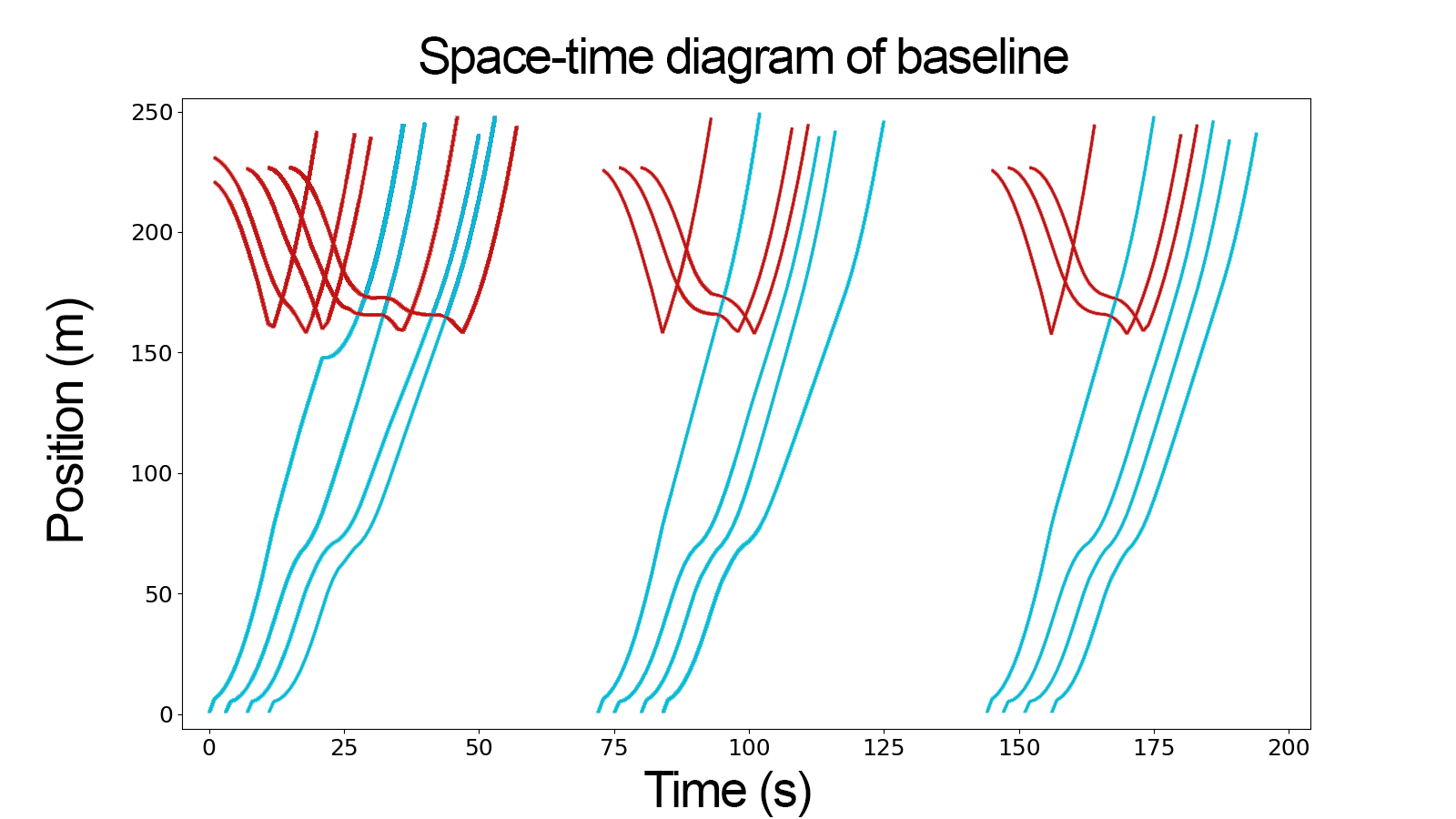}
    \includegraphics[width=0.45\textwidth]{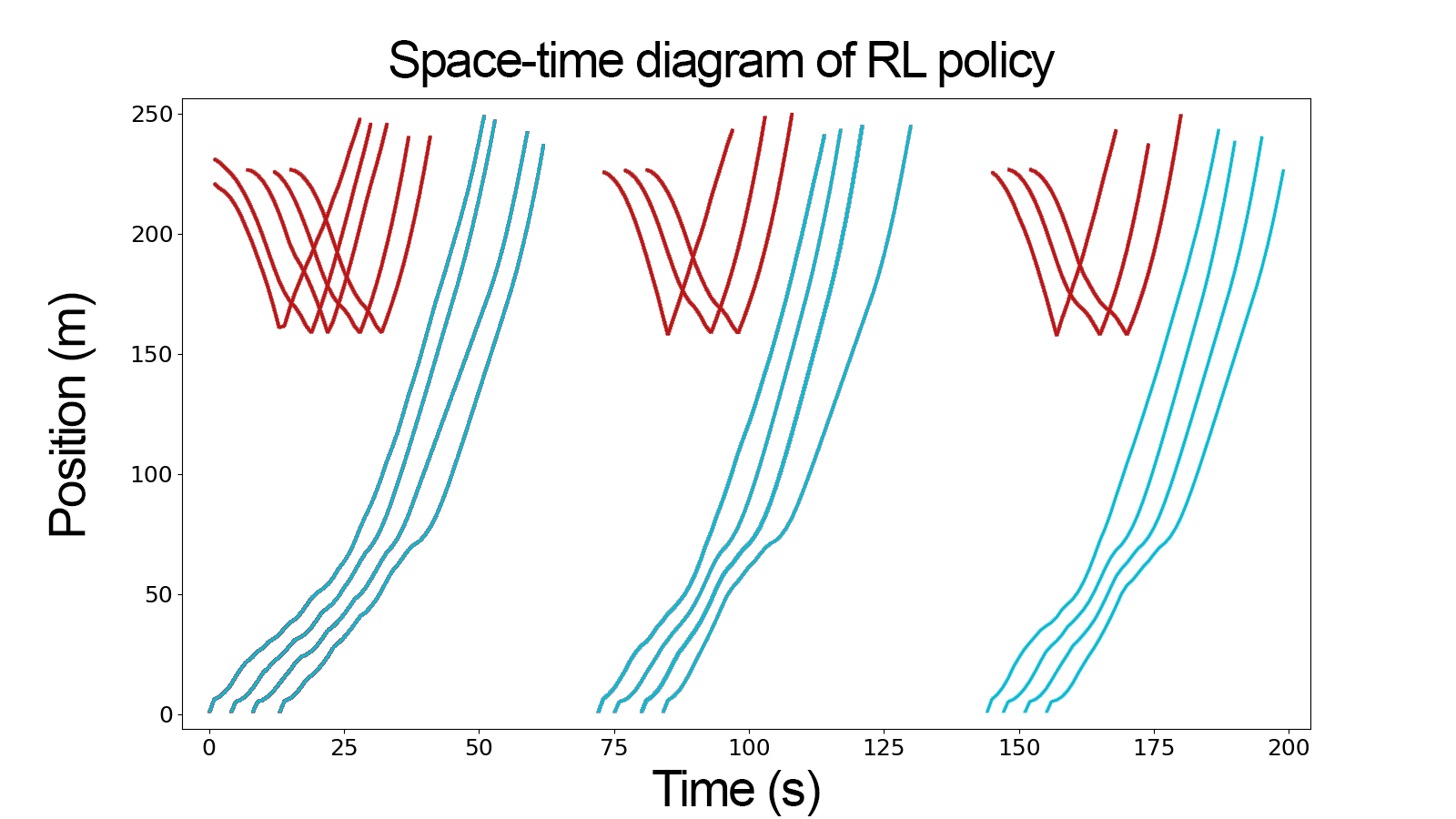}
    \caption{Space-time diagrams of the simulated baseline and RL policy. Each line corresponds to a vehicle in the system. Top: a guide to the color-scheme of the space time diagrams. The northern route is in red, the western route in blue. Middle: illustrates the overlap between the merging northern platoon and the western platoon. Bottom: The RL policy, depicted at the bottom successfully removes this overlap. Videos of this policy in simulation are available at \textcolor{blue}{\url{https://sites.google.com/view/iccps-policy-transfer}}.}
    \label{fig:space-time}
\end{figure}

\begin{figure}
\centering
\includegraphics[width=0.45\textwidth]{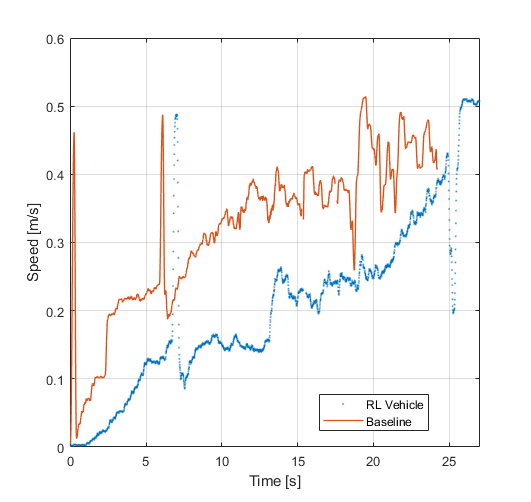}
\caption{Comparison of the first vehicle on the southern loop for the baseline (IDM) and RL experiments. The RL vehicle starts off slower but eventually accelerates sharply once the northern platoon has passed.}
\label{fig:velocity_profile}
\end{figure}


\begin{figure}
\centering
\includegraphics[width=0.45\textwidth]{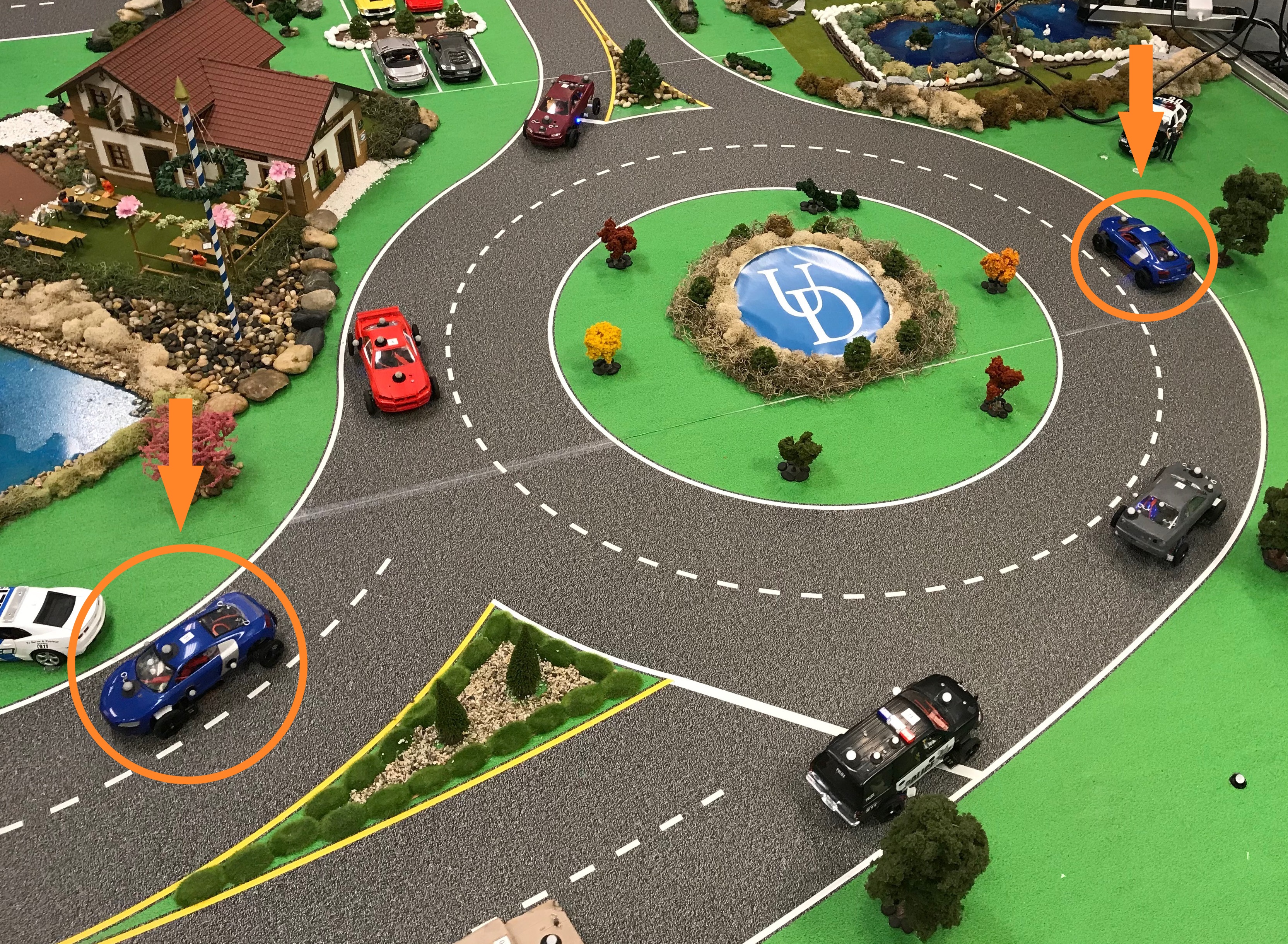}
\caption{Experiment with two platoons being led by RL vehicles (blue, circled).}
\label{fig:rl_UDSSC}
\end{figure}

In this section we present our results from a) training vehicular control policies via RL in simulation, and b) transferring the successful policy to UDSSC. Extended work and videos of the policies in action are available at \textcolor{blue}{\url{https://sites.google.com/view/iccps-policy-transfer}}.

\subsubsection{Simulation results}
Fig.~\ref{fig:reward} depicts the reward curve. The noise-injected RL policy takes longer to train than the noise-free policy and fares much worse during initial training, but converges to an almost identical final reward. In the simulations, videos of which are on the website, a ramp metering behavior emerges in which the incoming western vehicle learns to slow down to allow the vehicles on the north ramp to smoothly merge. 

This ramp metering behavior can also be seen in the space-time diagrams in Fig.~\ref{fig:space-time}, which portrays the vehicle trajectories and velocities of each vehicle in the system. Western vehicles are depicted in blue and northern in red. Due to the overlapping routes, visible in Fig.~\ref{fig:UDSSC-loop}, it was necessary to put a kink in the diagram for purposes of clarity; the kink is at the point where the northern and western routes meet. As can be seen in the middle figure, in the baseline case the two routes conflict as the northern vehicles aggressively merge onto the ramp and cut off the western platoon. Once the RL policy controls the autonomous vehicles, it slows down the western platoon so that no overlap occurs and the merge conflict is removed. 

\begin{figure*}[t]
\begin{multicols}{3}
  \includegraphics[width=\linewidth]{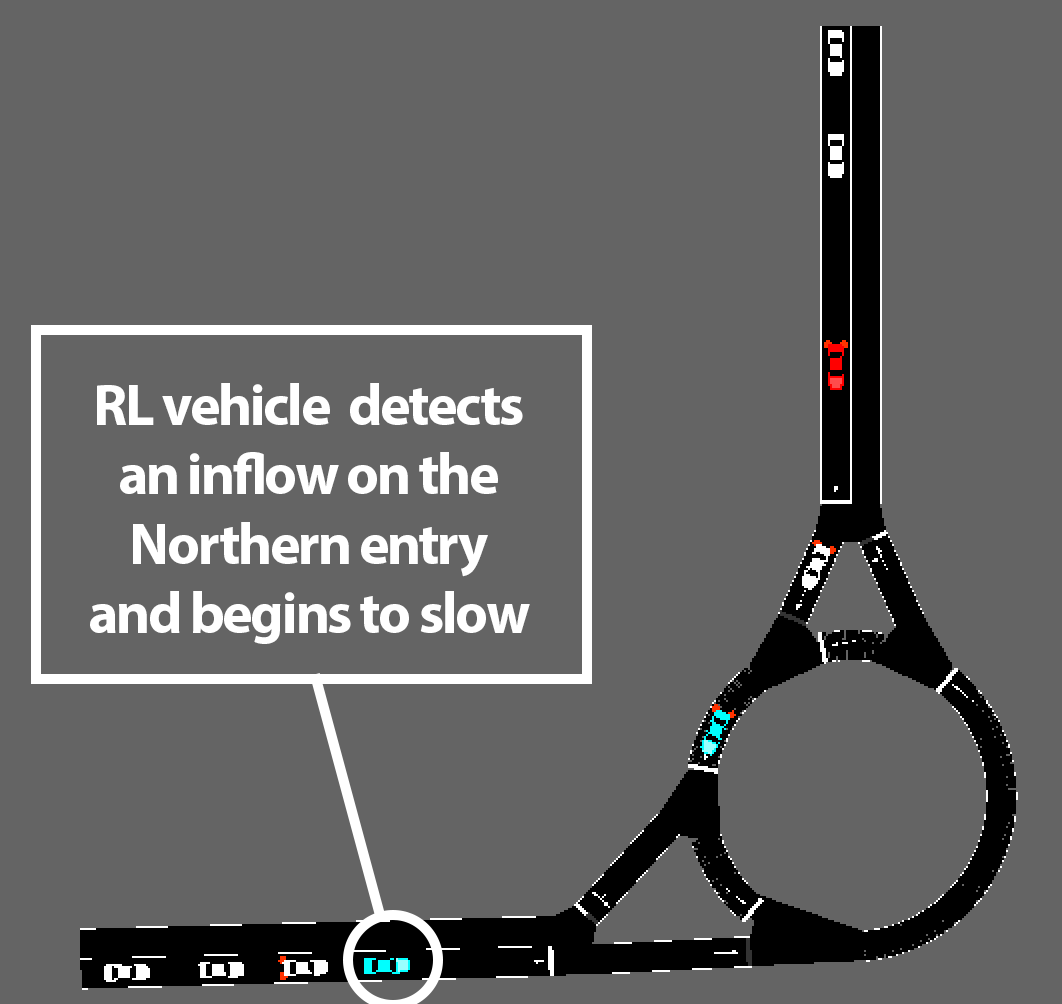} \par 
  \includegraphics[width=\linewidth]{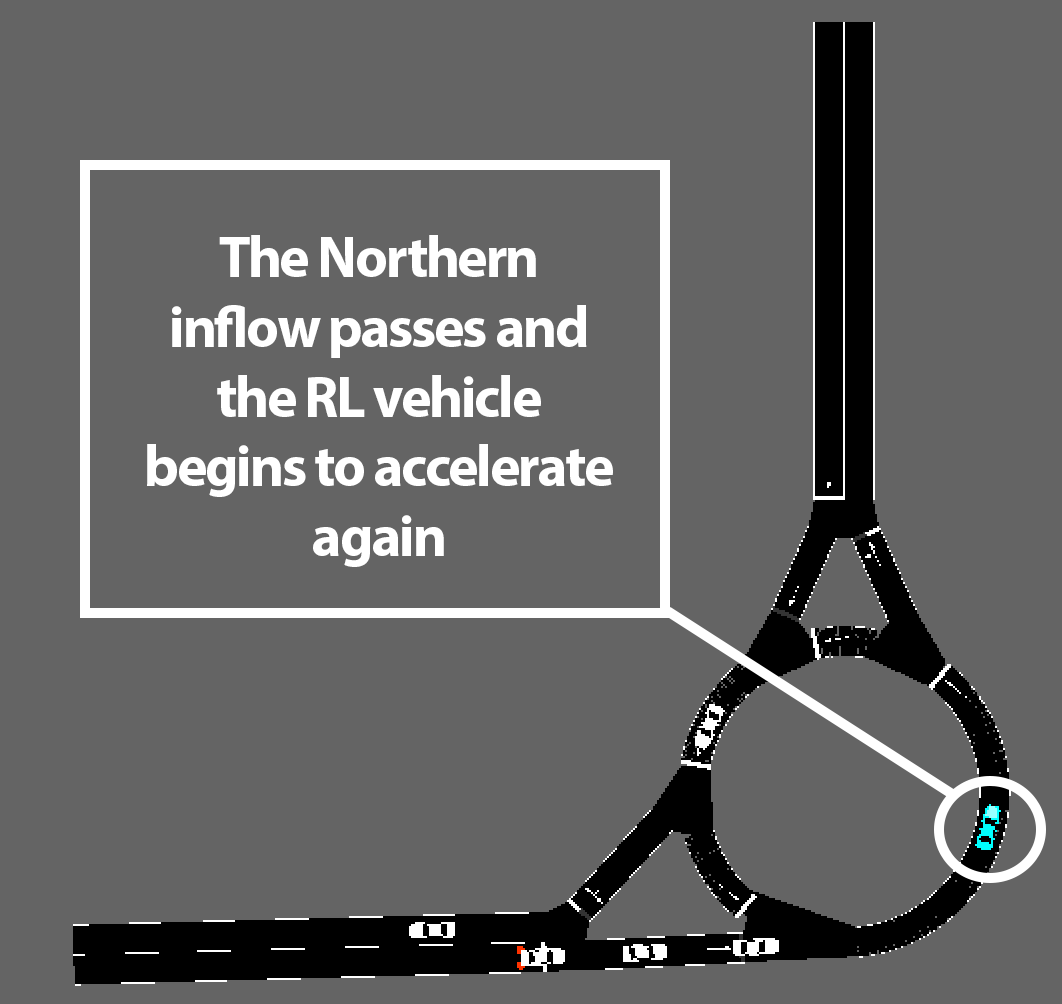}\par 
  \includegraphics[width=\linewidth]{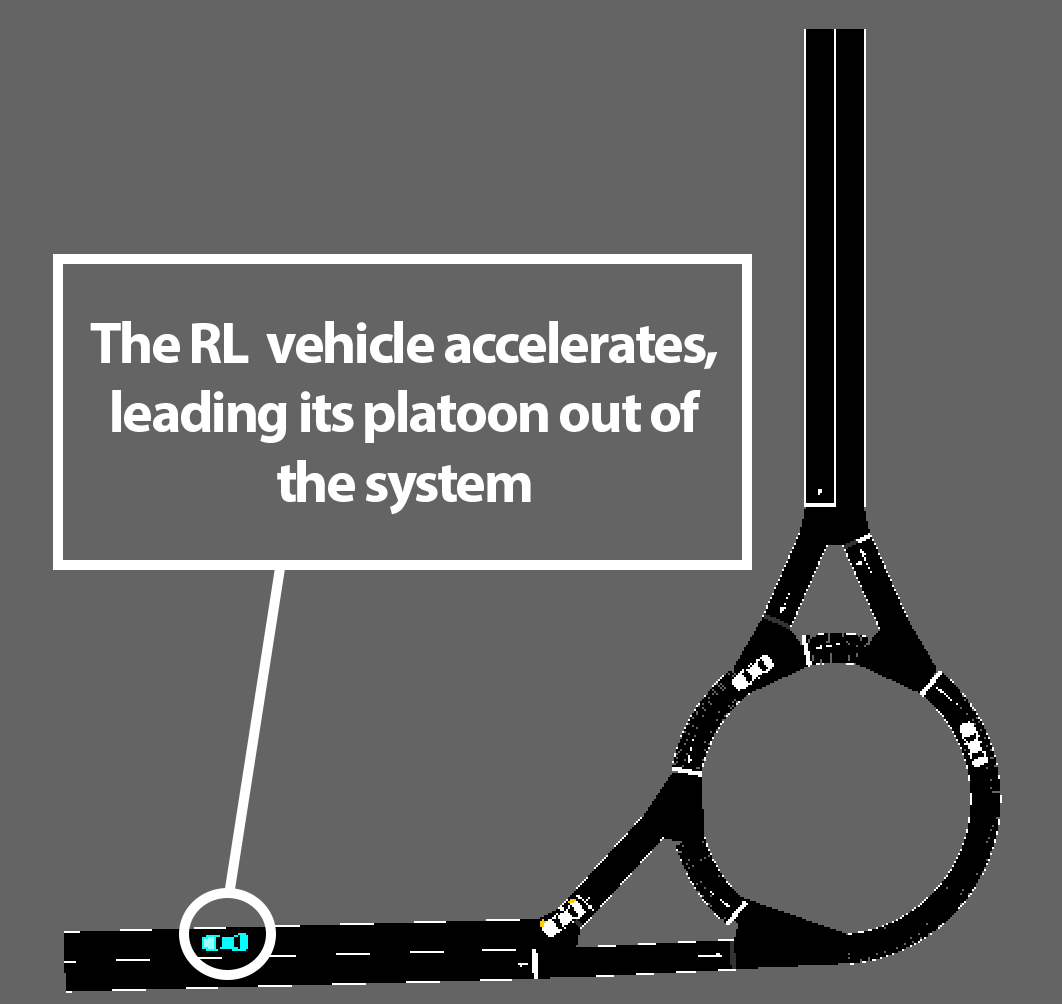} \par 
\end{multicols}
\caption{\footnotesize RL-controlled vehicle demonstrating smoothing behavior in this series of images. \textbf{First}: RL vehicle slows down in anticipation of a sufficiently short inflow from the north. \textbf{Second}: The northern inflow passes through the roundabout at high velocity. \textbf{Fourth}:The RL vehicle accelerates and leads its platoon away from the roundabout. Videos of this policy in simulation are available at \textcolor{blue}{\url{https://sites.google.com/view/iccps-policy-transfer}}.}
\label{fig:roundabout-progression}
\end{figure*}

\subsubsection{Transfer to UDSSC}
The RL policies were tested under three cases in UDSSC: (1) the baseline case with only vehicles running with the IDM controller, (2) the case with a leader vehicle running with the RL policy trained in sumo without additional noise, and (3) the case where the leader vehicles running with the RL policy were trained with noise actively injected into their observations and accelerations. The outcomes of these trials are presented in Table \ref{tab:congestionResults}.

During the congestion experiment, the third case, in which noise was actively injected into the action and state space during training, successfully exhibits the expected behavior it demonstrated in simulation. In this RL controlled case, the transfer consistently showed successful ramp metering: the western platoon adopted a lower speed than the baseline IDM controller, as can be seen in the lower velocity of the RL vehicle in Fig. ~\ref{fig:velocity_profile}. This allowed the northern queue to merge before the western platoon arrived, increasing the overall throughput of the roundabout. This is closer to a socially optimal behavior, leading to a lower average travel time than the greedy behavior shown in the baseline scenario. No unexpected or dangerous driving behavior occurred. 

This is in comparison with the second case, a policy trained on noiseless observations. In the second case, undesirable and unexpected deployment behavior suggests problems with the transfer process. Collisions occurred, sometimes leading to pile-ups, and platooning would frequently be timed incorrectly, such that, for example, only part of the Western platoon makes it through the roundabout before the Northern platoon cuts the Western platoon off. This indicates that the noise-injected policy is robust to transfer and resistant to domain and model mismatch.

Furthermore, the platoons led by RL vehicles trained with injected noise outperformed the baseline and noise-free cases. The results of these experiments, averaged over three trials with the RL vehicles, are presented in Table~ \ref{tab:congestionResults}. This improvement was the outcome of a metering behavior learned by the western RL platoon leader. In the baseline case, the north and western platoons meet and lead to a merge conflict that slows the incoming western vehicles down. This sudden decrease in speed can be seen in the drop in velocity at 20 seconds of the baseline in Fig.~\ref{fig:velocity_profile}.


Fig. \ref{fig:rl_UDSSC} shows the experiment in progress, with the blue (circled) vehicles being RL vehicles trained under noisy conditions. The RL vehicle entering from the western (lower) entrance has performed it's metering behavior, allowing vehicles from the northern (upper) queue to pass into the roundabout before the western RL vehicle speeds up again. Videos of the emergent behavior can be found on the website.

\begin{table}
\begin{tabular}{cccc}
    \toprule
     &Avg. Vel.[m/s]&Avg. Time[s]&Max Time[s] \\
     \midrule
     Baseline&0.26&15.71&23.99 \\
     RL&0.22&15.68&20.62 \\
     RL with Noise&0.23&14.81&18.68 \\
     \bottomrule
\end{tabular}
\caption{Results for the congestion experiment, the average and maximum times are averaged between three RL trials and a single baseline trial.}
\label{tab:congestionResults}
\end{table}

\section{Discussion}
\label{sec:discussion}
For the simulated environment, the choice of reward function, specifically, using the L2-norm rather than the L1-norm, encourages a more stable, less sparse solution. This makes evaluating the success of policy transfer more straightforward.
Table~\ref{tab:congestionResults} reports the results of the UDSSC experiments on three metrics:
\begin{itemize}
    \item The average velocity of the vehicles in the system
    \item The average time spent in the system
    \item The maximum time that any vehicle spent in the system
\end{itemize}

Note, the system is defined as the entire area of the experiments, including the entrances to the roundabouts. The layer of uncertainty in the noise-injected policy aided with overcoming the domain and model mismatch between the simulation system and the UDSSC system.
Thus, the noised policy was able to successfully transfer from simulation to UDSSC and also improved the average travel time by 5\% and the maximum travel time by 22\%. The noise-free policy did not improve on the average travel time and only improved the maximum travel time by 14\%. Although we did not perform an ablation study to check whether both state space noise and action space noise were necessary, this does confirm that the randomization improved the policy transfer process. 

The UDSSC consistently reproduced the moderate metering behavior for the noise-injected policy, but did not do so for the policy that was trained without noise. As can be seen in the videos, the noise-free policy was not consistent and would only irregularly reproduce the desired behavior, or meter dramatically to the point that average travel time increased. Overall the noised policy significantly outperformed the noise-free version.

However, we caution that in our testing of the policy transfer process on the UDSSC, we performed a relatively limited test of the effectiveness of the policy. The vehicles were all lined up outside the system and then let loose; thus, the tests were mostly deterministic. Any randomness in the tests would be due solely to randomness in the dynamics of the UDSSC vehicles and stochasticity in the transferred policy. In training, the acceleration of IDM vehicles are noised and can account for some stochasticity in the initial distribution of vehicles on the UDSSC. However, the trained policy was not directly given this inflow distribution at train time, so this does correspond to a separation between train and test sets.

There may be several reasons why the noise and action injection may have allowed for a successful zero shot transfer. First, because the action is noisy, the learned policy will have to learn to account for \emph{model mismatch} in its dynamics: it cannot assume that the model is exactly the double-integrator that is used in the simulator. Subsequently, when the policy is transferred to an environment with both delay, friction, and mass, the policy sees the mismatch as just another form of noise and accounts for it successfully. The addition of state noise helps with domain randomization; although the observed state distributions of the simulator and the scaled city may not initially overlap, the addition of noise expands the volume of observed state space in the simulator which may cause the two state spaces to overlap. Finally, the addition of noise forces the policy to learn to appropriately filter noise, which may help in the noisier scaled city environment.





\section{CONCLUSIONS}
\label{sec:conclusions}

In this paper, we demonstrated the real-world relevance of deep RL AV controllers for traffic control by overcoming the  gap between simulation and the real world. Using RL policies, AVs in the UDSSC testbed successfully coordinated at a roundabout to ensure a smooth merge. We trained two policies, one in the presence of state and action space noise, and one without, demonstrating that the addition of noise led to a successful transfer of the emergent metering behavior, while the noise-free policy often over-metered, or failed, to meter at all. This implies that for non-vision based robotic systems with small action-dimension, small amounts of noise in state and action space may be sufficient for effective zero-shot policy transfer. As a side benefit, we also demonstrate that the emergent behavior leads to a reduction of $5\%$ in average travel time and $22\%$ max-travel time on the transferred network. 

Ongoing work includes characterizing this result more extensively, evaluating the effectiveness of the efficiency of the policy against a wide range of vehicle spacing, platoon sizes, and inflow rates. In this context, there are still several questions we hope to address, as for example:

\begin{itemize}
\item Are both state and action space noise needed for effective policy transfer?
\item What scale and type of noise is most helpful in making the policy transfer?
\item Would selective domain randomization yield a less lossy, more robust transfer?
\item Would adversarial noise lead to a more robust policy?
\item Can we theoretically characterize the types of noise that lead to zero-shot policy transfer?
\end{itemize}
Finally, we plan to generalize this result to more complex roundabouts including many lanes, many entrances, and the ability of vehicles to change lanes. We also plan to evaluate this method of noise-injected transfer on a variety of more complex scenarios, such as intersections using stochastic inflows of vehicles.

\begin{acks}

  This research was funded in part by an NSF Graduate Research Fellowship and in part by the Delaware Energy Institute (DEI). AWS credits and funding were provided by an Amazon Machine Learning Research award. 

  The authors would also like to thank the Ray Zayas and Ishtiaque Mahbub for their contributions to enhancing the UDSSC database.

\end{acks}

\bibliographystyle{ACM-Reference-Format} 
\bibliography{mybib.bib,UDSSC.bib}


\begin{thebibliography}{29}


\ifx \showCODEN    \undefined \def \showCODEN     #1{\unskip}     \fi
\ifx \showDOI      \undefined \def \showDOI       #1{#1}\fi
\ifx \showISBNx    \undefined \def \showISBNx     #1{\unskip}     \fi
\ifx \showISBNxiii \undefined \def \showISBNxiii  #1{\unskip}     \fi
\ifx \showISSN     \undefined \def \showISSN      #1{\unskip}     \fi
\ifx \showLCCN     \undefined \def \showLCCN      #1{\unskip}     \fi
\ifx \shownote     \undefined \def \shownote      #1{#1}          \fi
\ifx \showarticletitle \undefined \def \showarticletitle #1{#1}   \fi
\ifx \showURL      \undefined \def \showURL       {\relax}        \fi
\providecommand\bibfield[2]{#2}
\providecommand\bibinfo[2]{#2}
\providecommand\natexlab[1]{#1}
\providecommand\showeprint[2][]{arXiv:#2}

\bibitem[\protect\citeauthoryear{Belletti, Haziza, Gomes, and Bayen}{Belletti
  et~al\mbox{.}}{2017}]%
        {belletti2017expert}
\bibfield{author}{\bibinfo{person}{Francois Belletti}, \bibinfo{person}{Daniel
  Haziza}, \bibinfo{person}{Gabriel Gomes}, {and} \bibinfo{person}{Alexandre~M
  Bayen}.} \bibinfo{year}{2017}\natexlab{}.
\newblock \showarticletitle{Expert level control of ramp metering based on
  multi-task deep reinforcement learning}.
\newblock \bibinfo{journal}{\emph{IEEE Transactions on Intelligent
  Transportation Systems}} (\bibinfo{year}{2017}).
\newblock


\bibitem[\protect\citeauthoryear{Bellman}{Bellman}{1957}]%
        {bellman1957markovian}
\bibfield{author}{\bibinfo{person}{Richard Bellman}.}
  \bibinfo{year}{1957}\natexlab{}.
\newblock \showarticletitle{A Markovian decision process}.
\newblock \bibinfo{journal}{\emph{Journal of Mathematics and Mechanics}}
  (\bibinfo{year}{1957}), \bibinfo{pages}{679--684}.
\newblock


\bibitem[\protect\citeauthoryear{Christiano, Shah, Mordatch, Schneider,
  Blackwell, Tobin, Abbeel, and Zaremba}{Christiano et~al\mbox{.}}{2016}]%
        {christiano2016transfer}
\bibfield{author}{\bibinfo{person}{Paul Christiano}, \bibinfo{person}{Zain
  Shah}, \bibinfo{person}{Igor Mordatch}, \bibinfo{person}{Jonas Schneider},
  \bibinfo{person}{Trevor Blackwell}, \bibinfo{person}{Joshua Tobin},
  \bibinfo{person}{Pieter Abbeel}, {and} \bibinfo{person}{Wojciech Zaremba}.}
  \bibinfo{year}{2016}\natexlab{}.
\newblock \showarticletitle{Transfer from simulation to real world through
  learning deep inverse dynamics model}.
\newblock \bibinfo{journal}{\emph{arXiv preprint arXiv:1610.03518}}
  (\bibinfo{year}{2016}).
\newblock


\bibitem[\protect\citeauthoryear{Cui, Seibold, Stern, and Work}{Cui
  et~al\mbox{.}}{2017}]%
        {cui2017stabilizing}
\bibfield{author}{\bibinfo{person}{Shumo Cui}, \bibinfo{person}{Benjamin
  Seibold}, \bibinfo{person}{Raphael Stern}, {and} \bibinfo{person}{Daniel~B
  Work}.} \bibinfo{year}{2017}\natexlab{}.
\newblock \showarticletitle{Stabilizing traffic flow via a single autonomous
  vehicle: Possibilities and limitations}. In
  \bibinfo{booktitle}{\emph{Intelligent Vehicles Symposium (IV), 2017 IEEE}}.
  IEEE, \bibinfo{pages}{1336--1341}.
\newblock


\bibitem[\protect\citeauthoryear{DOT}{DOT}{2016}]%
        {US2016}
\bibfield{author}{\bibinfo{person}{US DOT}.} \bibinfo{year}{2016}\natexlab{}.
\newblock \showarticletitle{National transportation statistics}.
\newblock \bibinfo{journal}{\emph{Bureau of Transportation Statistics,
  Washington, DC}} (\bibinfo{year}{2016}).
\newblock


\bibitem[\protect\citeauthoryear{Duan, Chen, Houthooft, Schulman, and
  Abbeel}{Duan et~al\mbox{.}}{2016}]%
        {duan2016benchmarking}
\bibfield{author}{\bibinfo{person}{Yan Duan}, \bibinfo{person}{Xi Chen},
  \bibinfo{person}{Rein Houthooft}, \bibinfo{person}{John Schulman}, {and}
  \bibinfo{person}{Pieter Abbeel}.} \bibinfo{year}{2016}\natexlab{}.
\newblock \showarticletitle{Benchmarking deep reinforcement learning for
  continuous control}. In \bibinfo{booktitle}{\emph{International Conference on
  Machine Learning}}. \bibinfo{pages}{1329--1338}.
\newblock


\bibitem[\protect\citeauthoryear{Gu, Holly, Lillicrap, and Levine}{Gu
  et~al\mbox{.}}{2017}]%
        {gu2017deep}
\bibfield{author}{\bibinfo{person}{Shixiang Gu}, \bibinfo{person}{Ethan Holly},
  \bibinfo{person}{Timothy Lillicrap}, {and} \bibinfo{person}{Sergey Levine}.}
  \bibinfo{year}{2017}\natexlab{}.
\newblock \showarticletitle{Deep reinforcement learning for robotic
  manipulation with asynchronous off-policy updates}. In
  \bibinfo{booktitle}{\emph{Robotics and Automation (ICRA), 2017 IEEE
  International Conference on}}. IEEE, \bibinfo{pages}{3389--3396}.
\newblock


\bibitem[\protect\citeauthoryear{Krajzewicz, Erdmann, Behrisch, and
  Bieker}{Krajzewicz et~al\mbox{.}}{2012}]%
        {SUMO2012}
\bibfield{author}{\bibinfo{person}{Daniel Krajzewicz}, \bibinfo{person}{Jakob
  Erdmann}, \bibinfo{person}{Michael Behrisch}, {and} \bibinfo{person}{Laura
  Bieker}.} \bibinfo{year}{2012}\natexlab{}.
\newblock \showarticletitle{Recent Development and Applications of {SUMO -
  Simulation of Urban MObility}}.
\newblock \bibinfo{journal}{\emph{International Journal On Advances in Systems
  and Measurements}} \bibinfo{volume}{5}, \bibinfo{number}{3\&4}
  (\bibinfo{date}{December} \bibinfo{year}{2012}), \bibinfo{pages}{128--138}.
\newblock


\bibitem[\protect\citeauthoryear{Kuindersma, Deits, Fallon, Valenzuela, Dai,
  Permenter, Koolen, Marion, and Tedrake}{Kuindersma et~al\mbox{.}}{2016}]%
        {kuindersma2016optimization}
\bibfield{author}{\bibinfo{person}{Scott Kuindersma}, \bibinfo{person}{Robin
  Deits}, \bibinfo{person}{Maurice Fallon}, \bibinfo{person}{Andr{\'e}s
  Valenzuela}, \bibinfo{person}{Hongkai Dai}, \bibinfo{person}{Frank
  Permenter}, \bibinfo{person}{Twan Koolen}, \bibinfo{person}{Pat Marion},
  {and} \bibinfo{person}{Russ Tedrake}.} \bibinfo{year}{2016}\natexlab{}.
\newblock \showarticletitle{Optimization-based locomotion planning, estimation,
  and control design for the atlas humanoid robot}.
\newblock \bibinfo{journal}{\emph{Autonomous Robots}} \bibinfo{volume}{40},
  \bibinfo{number}{3} (\bibinfo{year}{2016}), \bibinfo{pages}{429--455}.
\newblock


\bibitem[\protect\citeauthoryear{Levine and Abbeel}{Levine and Abbeel}{2014}]%
        {levine2014learning}
\bibfield{author}{\bibinfo{person}{Sergey Levine} {and} \bibinfo{person}{Pieter
  Abbeel}.} \bibinfo{year}{2014}\natexlab{}.
\newblock \showarticletitle{Learning neural network policies with guided policy
  search under unknown dynamics}. In \bibinfo{booktitle}{\emph{Advances in
  Neural Information Processing Systems}}. \bibinfo{pages}{1071--1079}.
\newblock


\bibitem[\protect\citeauthoryear{Li, Lv, and Wang}{Li et~al\mbox{.}}{2016}]%
        {li2016traffic}
\bibfield{author}{\bibinfo{person}{Li Li}, \bibinfo{person}{Yisheng Lv}, {and}
  \bibinfo{person}{Fei-Yue Wang}.} \bibinfo{year}{2016}\natexlab{}.
\newblock \showarticletitle{Traffic signal timing via deep reinforcement
  learning}.
\newblock \bibinfo{journal}{\emph{IEEE/CAA Journal of Automatica Sinica}}
  \bibinfo{volume}{3}, \bibinfo{number}{3} (\bibinfo{year}{2016}),
  \bibinfo{pages}{247--254}.
\newblock


\bibitem[\protect\citeauthoryear{Liang, Liaw, Nishihara, Moritz, Fox, Gonzalez,
  Goldberg, and Stoica}{Liang et~al\mbox{.}}{2017}]%
        {liang2017ray}
\bibfield{author}{\bibinfo{person}{Eric Liang}, \bibinfo{person}{Richard Liaw},
  \bibinfo{person}{Robert Nishihara}, \bibinfo{person}{Philipp Moritz},
  \bibinfo{person}{Roy Fox}, \bibinfo{person}{Joseph Gonzalez},
  \bibinfo{person}{Ken Goldberg}, {and} \bibinfo{person}{Ion Stoica}.}
  \bibinfo{year}{2017}\natexlab{}.
\newblock \showarticletitle{Ray RLLib: A Composable and Scalable Reinforcement
  Learning Library}.
\newblock \bibinfo{journal}{\emph{arXiv preprint arXiv:1712.09381}}
  (\bibinfo{year}{2017}).
\newblock


\bibitem[\protect\citeauthoryear{Mueller, Dosovitskiy, Ghanem, and
  Koltun}{Mueller et~al\mbox{.}}{2018}]%
        {muller2018transfer}
\bibfield{author}{\bibinfo{person}{Matthias Mueller}, \bibinfo{person}{Alexey
  Dosovitskiy}, \bibinfo{person}{Bernard Ghanem}, {and}
  \bibinfo{person}{Vladlen Koltun}.} \bibinfo{year}{2018}\natexlab{}.
\newblock \showarticletitle{Driving Policy Transfer via Modularity and
  Abstraction}. In \bibinfo{booktitle}{\emph{Conference on Robot Learning}}.
  \bibinfo{publisher}{IEEE}, \bibinfo{pages}{1--15}.
\newblock


\bibitem[\protect\citeauthoryear{Orosz}{Orosz}{2016}]%
        {orosz2016connected}
\bibfield{author}{\bibinfo{person}{G{\'a}bor Orosz}.}
  \bibinfo{year}{2016}\natexlab{}.
\newblock \showarticletitle{Connected cruise control: modelling, delay effects,
  and nonlinear behaviour}.
\newblock \bibinfo{journal}{\emph{Vehicle System Dynamics}}
  \bibinfo{volume}{54}, \bibinfo{number}{8} (\bibinfo{year}{2016}),
  \bibinfo{pages}{1147--1176}.
\newblock


\bibitem[\protect\citeauthoryear{Peng, Andrychowicz, Zaremba, and Abbeel}{Peng
  et~al\mbox{.}}{2017}]%
        {peng2017sim}
\bibfield{author}{\bibinfo{person}{Xue~Bin Peng}, \bibinfo{person}{Marcin
  Andrychowicz}, \bibinfo{person}{Wojciech Zaremba}, {and}
  \bibinfo{person}{Pieter Abbeel}.} \bibinfo{year}{2017}\natexlab{}.
\newblock \showarticletitle{Sim-to-real transfer of robotic control with
  dynamics randomization}.
\newblock \bibinfo{journal}{\emph{arXiv preprint arXiv:1710.06537}}
  (\bibinfo{year}{2017}).
\newblock


\bibitem[\protect\citeauthoryear{Pinto, Davidson, Sukthankar, and Gupta}{Pinto
  et~al\mbox{.}}{2017}]%
        {pinto2017robust}
\bibfield{author}{\bibinfo{person}{Lerrel Pinto}, \bibinfo{person}{James
  Davidson}, \bibinfo{person}{Rahul Sukthankar}, {and} \bibinfo{person}{Abhinav
  Gupta}.} \bibinfo{year}{2017}\natexlab{}.
\newblock \showarticletitle{Robust adversarial reinforcement learning}.
\newblock \bibinfo{journal}{\emph{arXiv preprint arXiv:1703.02702}}
  (\bibinfo{year}{2017}).
\newblock


\bibitem[\protect\citeauthoryear{Rios-Torres and Malikopoulos}{Rios-Torres and
  Malikopoulos}{2017}]%
        {Rios-Torres2017}
\bibfield{author}{\bibinfo{person}{Jackeline Rios-Torres} {and}
  \bibinfo{person}{Andreas~A. Malikopoulos}.} \bibinfo{year}{2017}\natexlab{}.
\newblock \showarticletitle{{A Survey on the Coordination of Connected and
  Automated Vehicles at Intersections and Merging at Highway On-Ramps}}.
\newblock \bibinfo{journal}{\emph{IEEE Transactions on Intelligent
  Transportation Systems}} (\bibinfo{year}{2017}), \bibinfo{pages}{1066--1077}.
\newblock
\showISBNx{15249050}
\showISSN{15249050}


\bibitem[\protect\citeauthoryear{Rusu, Vecerik, Roth{\"o}rl, Heess, Pascanu,
  and Hadsell}{Rusu et~al\mbox{.}}{2016}]%
        {rusu2016sim}
\bibfield{author}{\bibinfo{person}{Andrei~A Rusu}, \bibinfo{person}{Matej
  Vecerik}, \bibinfo{person}{Thomas Roth{\"o}rl}, \bibinfo{person}{Nicolas
  Heess}, \bibinfo{person}{Razvan Pascanu}, {and} \bibinfo{person}{Raia
  Hadsell}.} \bibinfo{year}{2016}\natexlab{}.
\newblock \showarticletitle{Sim-to-real robot learning from pixels with
  progressive nets}.
\newblock \bibinfo{journal}{\emph{arXiv preprint arXiv:1610.04286}}
  (\bibinfo{year}{2016}).
\newblock


\bibitem[\protect\citeauthoryear{Schulman, Levine, Abbeel, Jordan, and
  Moritz}{Schulman et~al\mbox{.}}{2015}]%
        {schulman2015trust}
\bibfield{author}{\bibinfo{person}{John Schulman}, \bibinfo{person}{Sergey
  Levine}, \bibinfo{person}{Pieter Abbeel}, \bibinfo{person}{Michael Jordan},
  {and} \bibinfo{person}{Philipp Moritz}.} \bibinfo{year}{2015}\natexlab{}.
\newblock \showarticletitle{Trust region policy optimization}. In
  \bibinfo{booktitle}{\emph{International Conference on Machine Learning}}.
  \bibinfo{pages}{1889--1897}.
\newblock


\bibitem[\protect\citeauthoryear{Shladover}{Shladover}{2017}]%
        {shladover2017connected}
\bibfield{author}{\bibinfo{person}{Steven~E Shladover}.}
  \bibinfo{year}{2017}\natexlab{}.
\newblock \showarticletitle{Connected and Automated Vehicle Systems:
  Introduction and Overview}.
\newblock \bibinfo{journal}{\emph{Journal of Intelligent Transportation
  Systems}} \bibinfo{number}{just-accepted} (\bibinfo{year}{2017}),
  \bibinfo{pages}{00--00}.
\newblock


\bibitem[\protect\citeauthoryear{Silver, Schrittwieser, Simonyan, Antonoglou,
  Huang, Guez, Hubert, Baker, Lai, Bolton, et~al\mbox{.}}{Silver
  et~al\mbox{.}}{2017}]%
        {silver2017mastering}
\bibfield{author}{\bibinfo{person}{David Silver}, \bibinfo{person}{Julian
  Schrittwieser}, \bibinfo{person}{Karen Simonyan}, \bibinfo{person}{Ioannis
  Antonoglou}, \bibinfo{person}{Aja Huang}, \bibinfo{person}{Arthur Guez},
  \bibinfo{person}{Thomas Hubert}, \bibinfo{person}{Lucas Baker},
  \bibinfo{person}{Matthew Lai}, \bibinfo{person}{Adrian Bolton},
  {et~al\mbox{.}}} \bibinfo{year}{2017}\natexlab{}.
\newblock \showarticletitle{Mastering the game of Go without human knowledge}.
\newblock \bibinfo{journal}{\emph{Nature}} \bibinfo{volume}{550},
  \bibinfo{number}{7676} (\bibinfo{year}{2017}), \bibinfo{pages}{354}.
\newblock


\bibitem[\protect\citeauthoryear{Stager, Bhan, Malikopoulos, and Zhao}{Stager
  et~al\mbox{.}}{2018}]%
        {Stager2018}
\bibfield{author}{\bibinfo{person}{Adam Stager}, \bibinfo{person}{Luke Bhan},
  \bibinfo{person}{Andreas Malikopoulos}, {and} \bibinfo{person}{Liuhui Zhao}.}
  \bibinfo{year}{2018}\natexlab{}.
\newblock \showarticletitle{A Scaled Smart City for Experimental Validation of
  Connected and Automated Vehicles}.
\newblock  \bibinfo{volume}{51}, \bibinfo{number}{9} (\bibinfo{year}{2018}),
  \bibinfo{pages}{130 -- 135}.
\newblock
\showISSN{2405-8963}
\newblock
\shownote{15th IFAC Symposium on Control in Transportation Systems CTS 2018.}


\bibitem[\protect\citeauthoryear{Swaroop and Hedrick}{Swaroop and
  Hedrick}{1996}]%
        {swaroop1996string}
\bibfield{author}{\bibinfo{person}{D Swaroop} {and} \bibinfo{person}{J~Karl
  Hedrick}.} \bibinfo{year}{1996}\natexlab{}.
\newblock \showarticletitle{String stability of interconnected systems}.
\newblock \bibinfo{journal}{\emph{IEEE transactions on automatic control}}
  \bibinfo{volume}{41}, \bibinfo{number}{3} (\bibinfo{year}{1996}),
  \bibinfo{pages}{349--357}.
\newblock


\bibitem[\protect\citeauthoryear{Tobin, Fong, Ray, Schneider, Zaremba, and
  Abbeel}{Tobin et~al\mbox{.}}{2017}]%
        {tobin2017domain}
\bibfield{author}{\bibinfo{person}{Josh Tobin}, \bibinfo{person}{Rachel Fong},
  \bibinfo{person}{Alex Ray}, \bibinfo{person}{Jonas Schneider},
  \bibinfo{person}{Wojciech Zaremba}, {and} \bibinfo{person}{Pieter Abbeel}.}
  \bibinfo{year}{2017}\natexlab{}.
\newblock \showarticletitle{Domain randomization for transferring deep neural
  networks from simulation to the real world}. In
  \bibinfo{booktitle}{\emph{Intelligent Robots and Systems (IROS), 2017
  IEEE/RSJ International Conference on}}. IEEE, \bibinfo{pages}{23--30}.
\newblock


\bibitem[\protect\citeauthoryear{Treiber, Hennecke, and Helbing}{Treiber
  et~al\mbox{.}}{2000}]%
        {Treiber2000}
\bibfield{author}{\bibinfo{person}{Martin Treiber}, \bibinfo{person}{Ansgar
  Hennecke}, {and} \bibinfo{person}{Dirk Helbing}.}
  \bibinfo{year}{2000}\natexlab{}.
\newblock \showarticletitle{Congested traffic states in empirical observations
  and microscopic simulations}.
\newblock \bibinfo{journal}{\emph{Physical review E}} \bibinfo{volume}{62},
  \bibinfo{number}{2} (\bibinfo{year}{2000}), \bibinfo{pages}{1805}.
\newblock


\bibitem[\protect\citeauthoryear{Vinitsky, Kreidieh, Le~Flem, Kheterpal, Jang,
  Wu, Liaw, Liang, and Bayen}{Vinitsky et~al\mbox{.}}{2018}]%
        {vinitsky2018benchmarks}
\bibfield{author}{\bibinfo{person}{Eugene Vinitsky}, \bibinfo{person}{Aboudy
  Kreidieh}, \bibinfo{person}{Luc Le~Flem}, \bibinfo{person}{Nishant
  Kheterpal}, \bibinfo{person}{Kathy Jang}, \bibinfo{person}{Fangyu Wu},
  \bibinfo{person}{Richard Liaw}, \bibinfo{person}{Eric Liang}, {and}
  \bibinfo{person}{Alexandre~M Bayen}.} \bibinfo{year}{2018}\natexlab{}.
\newblock \showarticletitle{Benchmarks for reinforcement learning in
  mixed-autonomy traffic}. In \bibinfo{booktitle}{\emph{Conference on Robot
  Learning}}. \bibinfo{publisher}{IEEE}, \bibinfo{pages}{399--409}.
\newblock


\bibitem[\protect\citeauthoryear{Wadud, MacKenzie, and Leiby}{Wadud
  et~al\mbox{.}}{2016}]%
        {Wadud2016}
\bibfield{author}{\bibinfo{person}{Zia Wadud}, \bibinfo{person}{Don MacKenzie},
  {and} \bibinfo{person}{Paul Leiby}.} \bibinfo{year}{2016}\natexlab{}.
\newblock \showarticletitle{Help or hindrance? The travel, energy and carbon
  impacts of highly automated vehicles}.
\newblock \bibinfo{journal}{\emph{Transportation Research Part A: Policy and
  Practice}}  \bibinfo{volume}{86} (\bibinfo{year}{2016}),
  \bibinfo{pages}{1--18}.
\newblock


\bibitem[\protect\citeauthoryear{Wu, Kreidieh, Parvate, Vinitsky, and Bayen}{Wu
  et~al\mbox{.}}{2017}]%
        {wu2017flow}
\bibfield{author}{\bibinfo{person}{Cathy Wu}, \bibinfo{person}{Aboudy
  Kreidieh}, \bibinfo{person}{Kanaad Parvate}, \bibinfo{person}{Eugene
  Vinitsky}, {and} \bibinfo{person}{Alexandre~M Bayen}.}
  \bibinfo{year}{2017}\natexlab{}.
\newblock \showarticletitle{Flow: Architecture and Benchmarking for
  Reinforcement Learning in Traffic Control}.
\newblock \bibinfo{journal}{\emph{arXiv preprint arXiv:1710.05465}}
  (\bibinfo{year}{2017}).
\newblock


\bibitem[\protect\citeauthoryear{Xu, Tang, and Tomizuka}{Xu
  et~al\mbox{.}}{2018}]%
        {xu2018transfer}
\bibfield{author}{\bibinfo{person}{Zhuo Xu}, \bibinfo{person}{Chen Tang}, {and}
  \bibinfo{person}{Masayoshi Tomizuka}.} \bibinfo{year}{2018}\natexlab{}.
\newblock \showarticletitle{Zero-shot Deep Reinforcement Learning Driving
  Policy Transfer for Autonomous Vehicles based on Robust Control}. In
  \bibinfo{booktitle}{\emph{International Conference on Intelligent
  Transportation Systems}}. \bibinfo{publisher}{IEEE},
  \bibinfo{pages}{2865--2871}.
\newblock


\end{thebibliography}

\end{document}